\documentclass[aip, apl, amsmath,amssymb,reprint, twocolumn]{revtex4}
\usepackage{graphicx}
\usepackage{color}

\begin{document}

\title{Superconducting micro-resonator arrays with ideal frequency spacing and extremely low frequency collision rate}

\author{X. Liu$^1$}
\author{W. Guo$^{2,3}$\footnote{Electronic mail: weijie.guo@nist.gov}}
\author{Y. Wang$^1$}
\author{M. Dai$^3$}
\author{L. F. Wei$^{1,3,4}$}
\author{B. Dober$^2$}
\author{C. McKenney$^2$}
\author{G. C. Hilton$^2$}
\author{J. Hubmayr$^2$}
\author{J. E. Austermann$^2$}
\author{J. N. Ullom$^2$}
\author{J. Gao$^{2}$}
\author{M. R. Vissers$^2$}

\affiliation{
1) Quantum Optoelectronics Laboratory, School of Physical Science and Technology, Southwest Jiaotong University, Chengdu, 610031, China\\
2) National Institute of Standards and Technology, Boulder, CO 80305, USA\footnote{Contribution of the U.S. Government, not subject to copyright}\\
3) Information Quantum Technology Laboratory, School of Information Science and Technology,Southwest Jiaotong University, Chengdu, 610031, China\\
4) State Key Laboratory of Optoelectronic Materials and Technologies, School of Physics, Sun Yat-Sen University, Guangzhou 510275, China}
\date{\today}

\begin{abstract}
We present a wafer trimming technique for producing superconducting micro-resonator arrays with highly uniform frequency spacing.
With the light-emitting diode (LED) mapper technique demonstrated previously, we first map the measured resonance frequencies to the physical resonators.
%We first map the measured resonance frequencies to the physical resonators by using our previously reported light-emitting diode (LED) mapper technique.
Then, we fine-tune each resonator's frequency by lithographically trimming a small length, calculated from the deviation of the measured frequency from its design value, from the interdigitated capacitor. We demonstrate this technique on a 127-resonator array made of titanium-nitride (TiN) and show that the uniformity of frequency spacing is greatly improved. The array yield in terms of frequency collisions improves from $84~\%$ to $97~\%$, while the quality factors and noise properties are unaffected.
%Because the trimming scheme can correct both the $\approx 1.5\%$ frequency deviation due to the uncertainty in design parameter and the $\leq 3\%$ radially position-dependent frequency non-uniformity due to the $T_{c}$ non-uniformity across the wafer, the measured resonance frequency after trimming can hit its designed value within $\approx 0.45\%$ accuracy. More importantly,
The wafer trimming technique provides an easy-to-implement tool to improve the yield and multiplexing density of large resonator arrays, which is important for various applications in photon detection and quantum computing.

\end{abstract}
\maketitle

Superconducting micro-resonators are important for many applications such as photon detection~\cite{Day}, quantum-limited amplifiers~\cite{Manuel}, readout of superconducting qubits~\cite{Wallraff} and readout of nano-mechanical resonators~\cite{Will}. They are of particular interest for superconducting detector applications because they are simple to fabricate and a large array of detectors can be read out through microwave frequency-domain multiplexing, which significantly reduces the complexity and cost of cryogenic wirings and readout electronics. For example, microwave kinetic inductance detectors (MKIDs)~\cite{Day, Doyle,Baselmans} are low-temperature detectors based on high-quality factor (high-$Q$) superconducting resonators~\cite{Zmuidzinas}. A large MKID array with thousands of resonators (pixels) can be fabricated with a small number of photo-lithography steps and read out with only a pair of coaxial cables into the cryostat. After over a decade of development, MKIDs are now used in both astronomical instruments~\cite{Baselmans2,Hubmayr,Mazin,Moore} and non-astronomical applications~\cite{Rowe, Jiansong, Weijie}.

%at the wavelength from sub-mm~\cite{Baselmans2,Hubmayr}, near-infrared and visible~\cite{Mazin}, X-ray~\cite{Day} to gamma-ray~\cite{Moore}. Other non-astronomical applications such as passive terahertz video-camera~\cite{Rowe} and photon-number resolving devices~\cite{Jiansong, Weijie} have also been demonstrated recently.

The pixel counts of large MKID arrays in major MKID instruments deployed and in development are growing rapidly, from hundreds (BLAST-TNG~\cite{BLAST}, NIKA~\cite{Monfardini}, MUSIC~\cite{Sayers}) to thousands (ARCONS~\cite{Eyken}, A-MKID~\cite{A-MKID}, NIKA 2~\cite{NIKA2}, TOLTEC~\cite{TOLTEC}) of pixels per wafer.
%such as BLAST~\cite{BLAST}, NIKA~\cite{Monfardini} with hundreds of pixels, MUSIC~\cite{Sayers} with $576$ pixels, ARCONS with 2024 pixels~\cite{Eyken}, and A-MIKD~\cite{A-MKID} with 21600 pixels used for APES.
As the array size and multiplexing density increase, resonator frequency collisions have become a serious problem that limit the array yield. The resonance frequencies are inevitably shifted from their design values due to various factors, such as non-uniformity in the superconducting critical temperature ($T_{c}$), film thickness, and over-etch depth across the wafer. When two resonators are overlapping or too close to each other in frequency space, referred to as a frequency collision, they cannot be read out properly due to cross-talk, thus reducing the effective array yield. Here, the yield is defined to be the number of useful non-colliding pixels over the total number of pixels. This problem becomes more severe in larger arrays because more resonators must be placed in a given frequency bandwidth, requiring smaller spacing between adjacent resonance frequencies, resulting in more unwanted frequency collisions.

%\textcolor{red}{Need a transition sentence to TiN.}
There are several approaches to reduce frequency collisions to improve the yield for large MKID arrays. The first approach is to increase the quality factor $Q$ of the resonator. But the highest design $Q$ for an application is usually capped by the required detector bandwidth and the responsivity. Another approach is to improve the fabrication process in order to achieve better wafer uniformity. For example, at NIST we have developed proximity-coupled TiN/Ti/TiN trilayer and multilayer films for MKIDs, which have greatly reduced the $T_c$ non-uniformity from over $20~\%$ to less than $2~\%$ across a $76.2$~mm wafer~\cite{Mike2013a,Mike2013b} as compared to sub-stoichiometric TiN films~\cite{ARCONS}. As the size of the array and the required multiplexing density continue to grow, it becomes more and more challenging to reduce frequency collisions through further improvement in the film uniformity.

%As an ideal material for MKID, titanium nitride (TiN) has received great attention for its high kinetic inductance, high normal resistivity and low loss in the GHz frequency range~\cite{Leduc,Mike2010}. TiN is preferred also because of the $T_{c}$ tunability ($0< T_{c} < 5$ K) by adjusting the nitrogen concentration. In practice, $T_{c} \approx 1$ K is suitable for most detector applications. However, $T_{c}$ is highly sensitive to the nitrogen concentration at the desired $T_{c} \approx 1$ K~\cite{Leduc}, resulting in a large non-uniformity of $T_c$ across the wafer~\cite{Mike2012,Mike2013}, which shifts the resonance frequencies from their designed values and leads to frequency collision. Previous results~\cite{ARCONS} show that the pixel yield is only $\approx 70\%$ due to frequency collision. To create more uniform TiN films, NIST are now developing trilayer/mutilayer TiN techniques~\cite{Mike2013} and the $T_{c}$ uniformity has greatly improved... Besides, advanced film fabrication techniques such as using atomic layer deposition rather than the traditional sputtering method~\cite{ALD,Szypryt} or finding new material to replace TiN~\cite{Szypryt2016}, are under explorations.

In this letter, we propose an alternative easy-to-implement technique based on two successive rounds of design, fabrication and measurement to produce a final resonator array with ideal frequency spacing and an extremely low frequency collision rate. In the second round, the resonance frequencies are re-tuned by lithographically trimming the interdigitated capacitor (IDC) of each resonator, using the measured resonance frequency information from the first round. We demonstrate this technique on a 127-resonator array from a TiN/Ti/TiN multilayer film and show that the array yield improves from $84~\%$ to $97~\%$, while the resonator quality factors and noise properties remain unaffected.

%This trimming technique also enables us to modify the measured resonance frequency to hit its designed value within $\approx 0.45\%$ accuracy. Meanwhile the performance ($Q$, dark noise, etc) of the trimmed pixel remains unchanged. Our wafer trimming technique is easy to implement and may be applied to any large resonator arrays for photon detector, qubit and SQUID multiplexer.

%In our previous work~\cite{LEDmapper}, we have demonstrated a LED wafer mapper technique to unambiguously correspond the measured frequency to detector pixel, which allows us to know the deviation of the measured frequency from the corresponding designed frequency. We can then re-design and trim each pixel to correct this frequency deviation.

%Our results show that the trimmed resonators have a more ideal frequency spacing and the pixel yield increases significantly from $86\%$ to $98\%$. This trimming technique also enables us to modify the measured resonance frequency to hit its designed value within $\approx 0.45\%$ accuracy. Meanwhile the performance ($Q$, dark noise, etc) of the trimmed pixel remains unchanged. Our wafer trimming technique is easy to implement and may be applied to any large resonator arrays for photon detector, qubit and SQUID multiplexer.
%caused by the inaccurate parameters in design and the $T_{c}$ non-uniform of the wafer.

In this work, we study lumped-element kinetic inductance detectors (LEKIDs)~\cite{Doyle} consisting of an inductor and an IDC.
%The resonance frequency of each LEKID is given by
%\begin{equation}
%f=\frac{1}{2\pi\sqrt{L\tilde{C}N_\mathrm{IDC}}}\label{eqn:fLC},
%\end{equation}
%where $L$ is the total inductance of the inductive strip including both the geometric inductance $L_{g}$ and kinetic inductance $L_{k}$, $\tilde{C}$ is the capacitance per IDC finger, and $N_\mathrm{IDC}$ is the number of IDC fingers.
In our LEKID array, all the resonators are designed to have the same inductor and the unique frequency of each resonator is defined by varying the number of IDC fingers.
%We vary $N_\mathrm{IDC}$ of each resonator to define their resonance frequencies.  This is because that trimming off IDC is much more easier to do in the fabrication process, compared to adding IDC to the resonator. Besides, depositing new layer onto the wafer may produce unpredictable effects while the simple implementation of IDC cutting can help remain the rest of the resonator unchanged. After proper IDC trimming, the resonance frequency shows a more uniform distribution. Fig.~1(a) shows the flow chart of our trimming technique towards an ideal frequency spacing and the detailed steps are listed as below:

The trimming technique for producing arrays with highly uniform frequency spacing consist of 6 steps which involve two rounds of design, fabrication and measurement. Fig.~1(a) shows a flow chart of the steps which are explained as below:

\begin{figure}[ht]
\includegraphics[width=8cm]{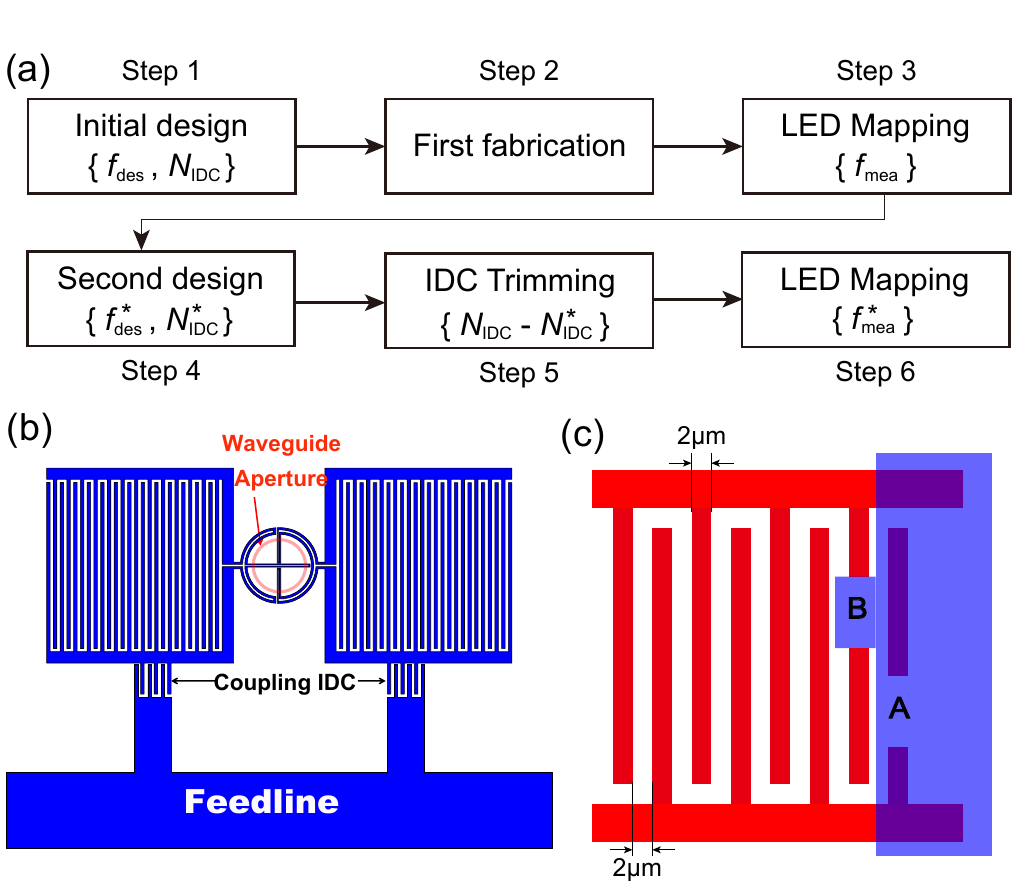}
%\includegraphics[width=8cm]{figure/sapceratio.pdf}
%\caption{Flow chart of trimming technique.}
\caption{(a) A flow chart of the trimming technique consisting of 6 steps with two rounds of design, fabrication and measurement. (b) Schematic drawing (not to scale) of the dual-polarization sensitive MKID single-pixel design~\cite{Dober}. %showing orthogonal X and Y polarization sensitive TiN absorbers attached to a pair of lumped-element MKIDs.  The waveguide aperture, which illuminates the inductive parts is depicted by the red shaded circular region.
(c) Schematic illustration of the IDC trimming process. Red area represents the remaining IDC fingers (2 $\mu$m finger/gap width). The blue area represents the area defined by a finger-trimming mask where TiN is etched off and IDC fingers are trimmed. In the first fabrication step (Step 2 in the flow chart), the first finger from the right was cut at position ``A" leaving a gap there. In the IDC trimming step (Step 5), the first finger from the right is entirely etched off and the second finger from the right was cut at position ``B". The total number of IDC fingers is reduced and the frequency is tuned upwards.}
\end{figure}

\textbf{Step 1: Initial design of the MKID array. }
%Along the winding direction of the feed-line, a certain resonator pixel (see fig.~1(b)) is labeled as number $i$ ($i$ = 1, 2, ...) and designed with a desirable frequency $f_\mathrm{des,i}$.
Assume we want to design an array of $N$ resonators with design frequencies of $f_\mathrm{des,i}$ ($i$ = 1, 2, .., $N$). The set $\{f_\mathrm{des,i}\}$ usually forms a frequency comb which fits within a certain readout bandwidth. We can determine the desired number of IDC fingers of each resonator $N_\mathrm{IDC,i}$ from the relation $f_\mathrm{des,i} =1/(2\pi\sqrt{L\tilde{C}N_\mathrm{IDC,i}})$, where the total inductance $L$ and capacitance per finger $\tilde{C}$ are values derived from electro-magnetic simulations and material parameters obtained from previous experiments. In this initial design step, we have ignored any wafer non-uniformity and assumed common $L$ and $\tilde{C}$ among all the resonators.

\textbf{Step 2: First fabrication.} In this step, the wafer is patterned into an array of resonators with the design number of IDC fingers $N_\mathrm{IDC,i}$.

\textbf{Step 3: Frequency measurement and LED mapping.} It is straightforward to measure the transmission $S_{21}$ of the entire array which contains all the resonances using a vector network analyzer (VNA). However, it is difficult to correctly correspond each resonance to its physical resonator on the wafer because the resonances are shifted from their design values due to wafer non-uniformity. In this step, we use the LED wafer mapping tool that we have previously developed especially for this purpose~\cite{LEDmapper} to establish the correspondence between the measured frequency $f_\mathrm{mea, i}$ and the number of fingers $N_\mathrm{IDC,i}$ of the $i$-th resonator on the wafer.

%We can easily correspond the resonator pixel $i$ to its measured resonance frequency $f_\mathrm{mea,i}$ by using the LED mapper technique. It is clear that the measured frequency $f_\mathrm{mea,i}$ deviates from its designed value $f_\mathrm{des,i}$ because of the uncertainty in $LC$ used in step 1. We can then retrieve the experimental value of $L_i\tilde{C}_i$ for each resonator from the relation $f_\mathrm{mea,i}=1/(2\pi\sqrt{L_i\tilde{C}_iN_\mathrm{IDC,i}})$. Note that the product of $L_i\tilde{C}_i$ is resonator dependent. $\tilde{C}_i$ should be the same for all the resonators because the IDC capacitance is basically independent of $T_{c}$. However, for resonator $i$, the total inductance $L_i$ is the combination of geometrical inductance $L_\mathrm{g,i}$ and kinetic inductance $L_\mathrm{k,i}$. $L_\mathrm{g,i}$ is $T_{c}$ independent but $L_\mathrm{k,i}$ is strongly dependent on $T_{c}$ through the relation $L_{k}\propto \hbar R_\mathrm{sn}/T_{c}$~\cite{Mattis1958,Gao2014}. Therefore a lower $T_{c}$ translates into a larger kinetic inductance. Because of the non-uniformity of $T_c$ across the 75 mm wafer, $L_i\tilde{C}_i$ is position dependent.\\

\textbf{Step 4: Array re-design.}
 To account for the deviation of $f_\mathrm{mea,i}$ from $f_\mathrm{des,i}$ due to wafer non-uniformity, we set up a new model $f_\mathrm{mea,i}=1/(2\pi\sqrt{L_i\tilde{C}_iN_\mathrm{IDC,i}})$, where $L_i$ and $\tilde{C}_i$ are introduced to account for the local variations of inductance and capacitance at the location of the $i$-th resonator. By analyzing the $f_\mathrm{mea, i}$ vs. $N_\mathrm{IDC,i}$ data obtained from Step 3, we retrieve the local $L_i\tilde{C}_i$ product of each resonator. We then set up a new frequency comb $\{f^*_\mathrm{des, i}\}$ as our new design frequencies. We require that $f^*_\mathrm{des, i} \gtrsim f_\mathrm{mea, i}$, because in the next step we will etch off a small portion from the original IDC fingers. Calculated from  $f^*_\mathrm{des,i}=1/(2\pi\sqrt{L_i\tilde{C}_iN^*_\mathrm{IDC,i}})$ using the derived $L_i\tilde{C}_i$ values, the $i$-th resonator is re-designed to have a new number of IDC fingers $N^*_\mathrm{IDC,i}$ ($\le N_\mathrm{IDC,i}$).

\textbf{Step 5: IDC trimming.} In this step, the wafer is sent back to the clean room and a small portion of each IDC is trimmed off using lithographical tools and reactive ion etching which reduces the number of IDC fingers from $N_\mathrm{IDC,i}$ to $N^*_\mathrm{IDC,i}$.

\textbf{Step 6: Frequency re-measurement of the trimmed array.} In this step, we repeat the VNA sweep and LED mapping procedures in Step 2 to obtain the new resonance frequencies $\{f^*_\mathrm{mea,i}\}$. Comparison and statistics are carried out on both $f^*_\mathrm{mea,i}$ and $f^*_\mathrm{des,i}$ to evaluate the effect of the array trimming technique.

%We are developing feed-horn coupled dual-polarization sensitive MKID arrays~\cite{Hubmayr} made from titanium-nitride/titanium/titanium-nitride (TiN/Ti/TiN) trilayer films with target $T_{c}\approx 1.4$~K for the BLAST-TNG experiment~\cite{BLAST}.
%First design
To demonstrate our array trimming technique, we made a 127-pixel hexagonal close-packed LEKID array from a $60$ nm thick multilayer TiN/Ti/TiN film ($T_c \approx 1.6$~K) on a $76.2$~mm intrinsic Si wafer. Each pixel is identical to the BLAST-TNG pixel design~\cite{Dober} in the $500$ $\mu$m band, which consists of orthogonal TiN absorbers attached to a pair of IDC (Fig.~1(b)). In this letter we study only the $127$ X-pol resonators.
%First design

In the initial design step (Step 1), the design resonance frequencies are in a geometric series as $f_\mathrm{des,i} = 750$~MHz $\times 1.002^{i-1}$, where $i$ = 1,...,127 is the resonator index along the meandering feedline on the wafer. The coupling quality factor is designed to be $Q_c \approx 30,000$, which will yield a total quality factor $Q \approx 20,000$ under optical loading.
%First fab, stepper

In the first fabrication step (Step 2), we patterned the wafer using a semi-automated ``tiling and trimming" technique developed for BLAST-TNG arrays with a stepper tool~\cite{McKenney}. In this technique, a mask of the standard pixel tile (Fig.~1(b)) containing the common inductor part and an IDC with the maximum number of fingers, is repeatedly exposed onto the wafer at the desired pixel positions. A second set of IDC trimming masks (blue area in Fig.~1(c)) is used to cut each resonator's IDC fingers at the desired position (position ``A" in Fig.~1(c)) to create the designed frequency comb. This fabrication technique limits the number of stepper masks, ensures high quality and uniformity in the stepper lithography because all pixels are patterned from the same standard pixel mask, and provides considerable flexibility in resonator frequency definition and array layout.

%First mapping, s21, collision, LED
After the wafer is made, we cooled it down along with the LED-mapper setup in a dilution refrigerator. We measured the transmission $S_\mathrm{21}$ of the MKID array at the base temperature of $40$ mK using a VNA and all the resonances were matched to their physical pixel on the wafer using the LED-mapper tool. As shown by the red curve in Fig.~2, 127 resonance dips are clearly observed in the frequency range spanned from $700$~MHz to $930$~MHz. Although the resonators are designed to have ideally spaced frequencies, some are too close and collide with their neighbors. If we use a 5-linewidth criterion for the frequency collision (the spacing between adjacent resonances is smaller than $5 f/Q$), then $20$ resonances among the total $127$ resonances are in collision (taking $Q = 20,000$), resulting in a much reduced array yield of $84~\%$, despite the fabrication yield of $100~\%$.

%We know multilayer film is more uniform than the sub-stoichiometric TiN film, but the finite non-uniformity of $T_{c}$ still causes the frequencies shift from their designed values due to the inductance dependence on $T_{c}$~\cite{LEDmapper}. Therefore we develop a trimming technique towards an ideal frequency distribution without improving the TiN film uniformity.

\begin{figure}[h]
\includegraphics[width=8.5cm]{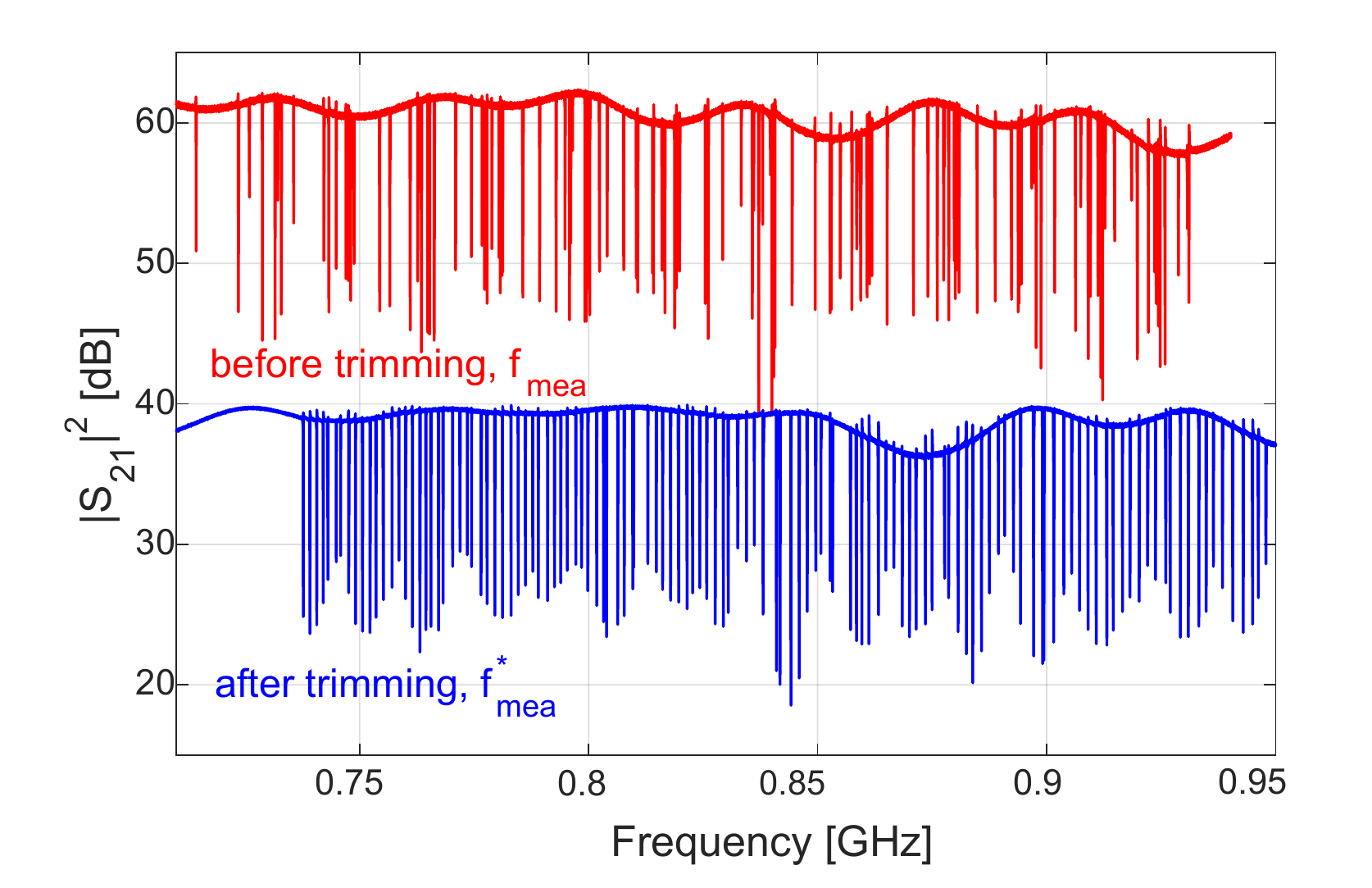}
\caption{A comparison of $S_\mathrm{21}$ for the MKID array before (red) and after (blue) trimming process. The red curve is moved up by $20$~dB for clarity. Note that the resonance depths are unchanged after trimming, suggesting that the quality factors are unchanged.}
\end{figure}
%First, we can use LED mapper technique~\cite{LEDmapper} to obtain the frequency to pixel-position correspondence data, i.e., we are able to know the measured frequency and designed frequency for each resonator. Then we can design a new set of ideally spaced frequencies $f^*_\mathrm{des,i}$ and determine the corresponding new set of IDC finger number $N^*_\mathrm{IDC,i}$ for each resonator.

%2nd design, fdes = . NIDC reduce, 76 stay
In the re-design step (Step 4), in order to correct the effect of wafer non-uniformity, we selected a new set of design frequencies $f^*_\mathrm{des,i} = 739$~MHz $\times 1.002^{i-1}$ and designed a new set of $N^*_\mathrm{IDC,i}$ based on the local $L_i\tilde{C}_i$ values obtained in Step 3.

%2nd fab
In the next re-fabrication step (Step 5), the wafer was sent back to the clean room and a small portion of IDC finger ($N_\mathrm{IDC,i}-N_\mathrm{IDC,i}^*$) was lithographically patterned and removed for each resonator $i$ that requires frequency trimming (Fig.~1(c)). While it is possible to use the stepper tool and the same IDC trimming mask again to cut the IDC fingers, we used a maskless aligner (MLA) tool which is faster and more convenient. It directly exposes a pattern containing all areas to be etched off from a file onto the entire wafer without the need for a chrome mask. Although the lithography resolution of the MLA ($\sim$1 micrometer) is poorer than the stepper (sub-micrometer), it is adequate for our purpose and the lithographical error has a negligible effect on the resonance frequency.

%2nd mapping, s21, LED, collision rate
After IDC trimming, we measured the MKID wafer at $40$~mK again and the $S_\mathrm{21}$ is shown by the blue curve in Fig.~2. One can see that the resonances are globally shifted up by about $20$~MHz since we have reduced the number of IDC fingers for most resonators. It is obvious that the uniformity of the frequency spacing after trimming is greatly improved. According to the 5-linewidth criterion of frequency collision, we find only $4$ colliding resonances out of a total of $127$ resonances, thus our technique has effectively improved the overall array yield from $84~\%$ to $97~\%$.

\begin{figure}[ht]
\includegraphics[width=8.5cm]{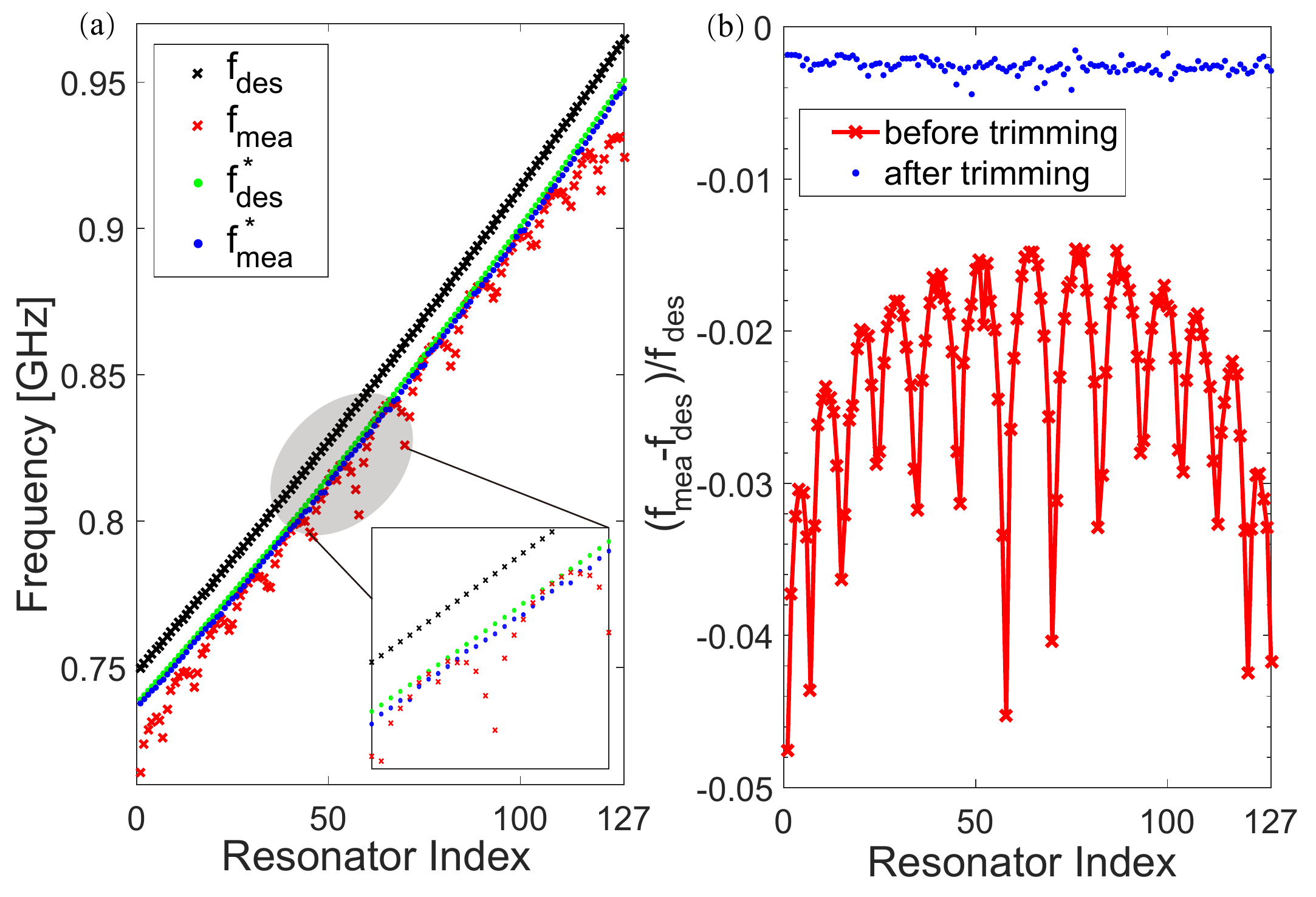}
\caption{(a) Design and measured resonance frequency vs. resonator (pixel) index. The black cross and red cross represent the design frequency ($f_\mathrm{des}$) and the corresponding measured frequency ($f_\mathrm{mea}$) before trimming. The green point and blue point are the design frequency $f_\mathrm{des}^*$ and measured frequency $f_\mathrm{mea}^*$ after trimming. Compared to the data before trimming, the periodic deviation vanishes and the deviation of $f_\mathrm{mea}^*$ from $f_\mathrm{des}^*$ is too small to distinguish in Fig.~3(a). (b) The fractional frequency deviation vs. resonator index number, which shows the trimming technique significantly reduces the deviation between the design and measured frequency.}
%(c) Color-coded surface plots visualizing the position-dependent $\delta$ on the 76.2~mm wafer before trimming. The black dots are the center positions of the measured pixels. The numbers in black represent the numerical order of the measured resonance frequencies from low to high, which are not always consistent with the resonator index number, i.e., frequency re-shuffling occurs although the frequency is designed to monotonically increase with resonator index number. (d) Color-coded surface plots visualizing the position-dependent $\delta$ on the 76.2~mm wafer after trimming. The measured resonance $f_\mathrm{mea}^*$ is now completely consistent with the resonator index number and the frequency deviation $\delta$ is about $1$ order smaller than before trimming.
\end{figure}

%We further investigated the effect of the trimming technique on the uniformity of resonance frequencies.
Fig.~3(a) shows the design and measured frequency before and after trimming.
The fractional frequency deviation $\delta=(f_\mathrm{mea}-f_\mathrm{des})/f_\mathrm{des}$, which reveals the quantitative agreement between the measured and design frequencies, is plotted in Fig.~3(b). Before trimming, the $f_\mathrm{mea}$ exhibit an obvious periodic deviation from $f_\mathrm{des}$ as shown in both Fig.~3(a) and (b). As discussed in our previous paper~\cite{LEDmapper}, this is caused by a radial non-uniformity of $T_c$ across the wafer in multilayer TiN/Ti/TiN films. Resonators near the turns of the meandering feedline are in general located on the outer ring of the wafer where the $T_c$ is lower and kinetic inductance is higher, leading to the larger negative deviation of the resonance frequency. On average, $f_\mathrm{mea}$ is lower by a factor of $\sim 2.4\times10^{-2}$ than its design value $f_\mathrm{des}$, which may be attributed to the inaccurate $L$ and $C$ values used in the initial design and the wafer non-uniformity. It is clear that the measured frequency is much closer to the design frequency after the trimming process, as shown in Fig.~3(b). The maximum deviation $|\delta|_\mathrm{max}$ is only $4.5\times10^{-3}$ which is significantly smaller than that before trimming. We notice that all the measured frequencies shifted down by $\sim 2\times10^{-3}$ from the design frequencies. As a reference, the $76$th resonator is untrimmed during the re-fabrication process ($N_\mathrm{IDC,76}=N^*_\mathrm{IDC,76}$) but its frequency also shifted down by $2\times10^{-3}$ like other trimmed resonators. This suggests that the global shift is a systematic effect (such as surface oxidation), which is not associated with the array trimming process.

%Ideally, we should get $f_\mathrm{mea,76}^{*}=f_\mathrm{mea,76}=f_\mathrm{des,76}^{*}$ and $\delta_\mathrm{76}^{*}=0$, i.e., the measured frequency of resonator $76$ should be the same as in the first measurement since this resonator is entirely unchanged.

%However the measured $\delta_\mathrm{76}^{*}=0.2\%$, indicating that the finite $\approx 0.2\%$ overall shift is due to the oxidation of wafer surface. Besides, we use LED mapper technique to correspond the trimmed resonators to their measured frequencies $f_\mathrm{mea}^*$ and find that $f_\mathrm{mea}^*$ now arranges in order, i.e., resonance frequency increases monotonically with resonator number, which is in good agreement with our design. In contrast, for the wafer before trimming, frequency re-shuffling occurs on a few adjacent resonators. These results are summarized in fig.~3(c) and fig.~3(d).

%According to the result from LED mapper paper ~\cite{LEDmapper}, we find our TiN/Ti/TiN multilayer film shows a $6\%$ variation of $T_{c}$ ($T_{c} \approx$ 1.4K) cross the wafer.

\begin{figure}[htb]
\includegraphics[width=8cm]{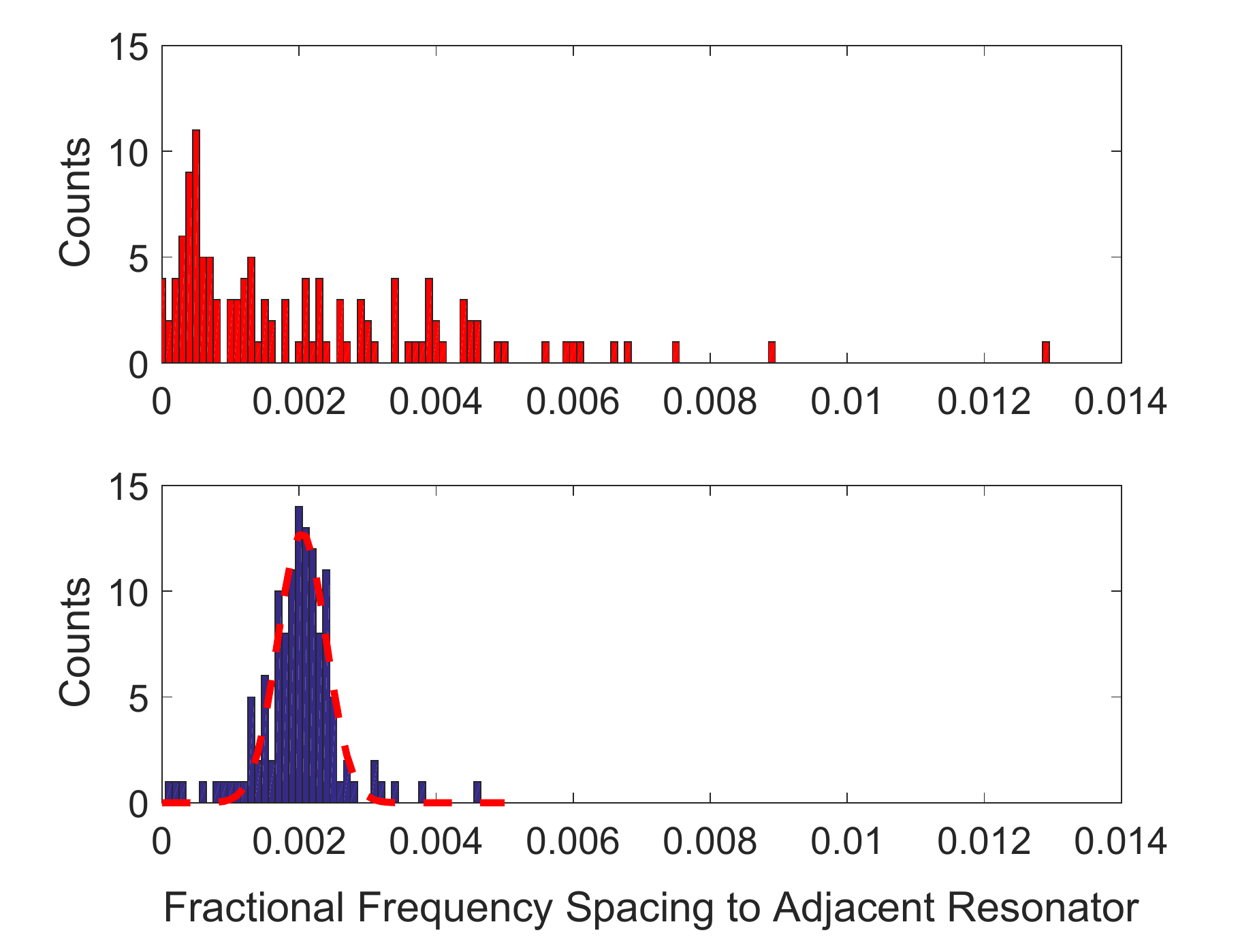}
\caption{(a) Histogram of the fractional frequency spacing before the trimming process. (b) Histogram of the fractional frequency spacing after the trimming process. The red dash line fits to a Gaussian distribution $f(x)=\frac{1}{\sqrt{2\pi}\sigma} e^{-\frac{(x-\mu)^2}{2\sigma^2}}$ with $\mu = 0.002$ and $\sigma = 3.5 \times 10^{-4}$.}
%before trimming:$1.72\pm 1.63$ MHz.
%after trimming:$1.67\pm0.52$ MHz.
%two standard deviations from the mean (medium and dark blue) account for 95.45 percent
\end{figure}

Fig.~4 compares the distribution of fractional frequency spacing for the MKID array before and after the trimming process. Here the fractional frequency spacing are calculated from $
\delta f/f =  (f_\mathrm{mea, n+1}-f_\mathrm{mea,n})/f_\mathrm{mea,n},~n = 1,...,126$. Before trimming, $\delta f/f$ shows larger scatter from $1 \times 10^{-5}$ to $1.3\times10^{-2}$ while the design fractional spacing is $2\times10^{-3}$. The scatter is greatly reduced after the trimming process. Taking $Q = 20,000$, the 5-linewidth frequency collision criterion excludes $20$ (or $16~\%$) resonators before trimming and only $4$ (or $3~\%$) resonators after trimming. Therefore, the array yield is significantly improved.

The histogram in Fig.~4(b) fits well to a Gaussian distribution, which allows us to predict the ultimate array yield from the derived formula ~\cite{Supl}
\begin{equation}
P_{0}=\{\prod_{n=1}^{n=\infty}[1-\frac{\mathrm{Erf}(\frac{n\Delta}{\sqrt{2}\sigma}+\frac{\chi w}{\sqrt{2}\sigma})-\mathrm{Erf}(\frac{n\Delta}{\sqrt{2}\sigma}-\frac{\chi w}{\sqrt{2}\sigma})}{2}]\}^2,
\end{equation}
%If we assuming a Gaussian distribution of frequency spacing, the ultimate array yield can be derived by considering the probability $P_{0}$ for a resonator in a frequency comb (with a large number of resonators both below and above its frequency) to survive the collisions with all the other resonators~\cite{Supl},
where $P_{0}$ is the probability for a resonator to survive collisions with all the other resonators, $w=1/Q$ is the normalized resonator linewidth, $\Delta$ is the designed fractional frequency spacing, and $\sigma$ is the standard deviation of the Gaussian distribution. We set the number of linewidths to exclude to $\chi=5$, which is the criterion used for frequency collision throughout this letter. As an example, the yield of a MKID array with $1000$ resonators distributed in an octave bandwidth is predicted to be $81~\%$. This yield and the multiplexing density are desirable for many MKID instruments and their readout electronics.
%MonteCarlo is 87%; Q=20K, yield is 96.7\%; 10K is 85\%.

Last, we have also verified that the quality factor and noise of the resonators remain unchanged before and after the trimming process. Therefore the detector performance is not affected.

%$df=f_{mea}(n+1)-f_{mea}(n)$. We can see the frequency spaces between two adjacent resonators before trimming are from $0.01$ MHz to $9.2$ MHz. There are $9$ frequency spaces smaller than $0.2$ MHz ($18$ resonators involved). As a comparison, the frequency spaces after trimming range from $0.07$MHz to $3.90$ MHz, and only $2$ frequency spaces below $0.2$ MHz (4 resonators involved). Most of frequency space center on the area from $1$ Hz to $2.5$MHz as wanted. Furthermore, the performance, including Q and noise, is still same as original.

In conclusion, we have demonstrated a wafer trimming technique which combines two successive rounds of design, fabrication and measurement together to produce a final resonator array with ideal frequency spacing and extremely low frequency collision rate. We use this technique on a 127-resonator array made from a TiN/Ti/TiN multilayer film and show that the array yield improves from $84~\%$ to $97~\%$, while the resonator quality factors and noise properties remain unaffected. We have also demonstrated that the measured resonance frequency matches its design value within an accuracy of $4.5\times10^{-3}$ after the trimming process.

The proposed wafer trimming technique and its principle are applicable to any superconducting films, even films with poor wafer uniformity, and other resonator types than lumped-element resonators. It can also be easily modified to re-adjust other parameters of the array (such as the coupling $Q_c$). Our technique provides an easy-to-implement and effective tool to improve the yield and multiplexing density of large resonator arrays, which may find broad applications in photon detection and quantum computing.

%It provides a simple and effective way to make the resonant frequency more uniform, although the $T_{c}$ across the wafer is still non-uniform. First, we use LED mapper scheme to correspond the measured resonant frequency to the pixel. And then, we can re-design and trim each pixel individually after having the relationship between the design frequency and measured frequency. The trimmed wafer shows ideally frequency-space distribution. And the detector yield which often decreases due to frequency collisions in the readout is improved from $86\%$ to $98\%$. And we can hit design resonance frequency accurately with $0.45\%$ by using LED trimming scheme. More important, it is easy to implement and the per-pixel performance is as good as the pixel before trimming. This simple and efficiency LED trimming scheme can be widely used in any large resonator array.

The MKID devices were fabricated in the NIST-Boulder microfabrication facility. We thank Prof. Xie Dong for his technical assistance. This work was supported in part by the National Natural Science Foundation of China (Grant Nos. 61301031, U1330201).

\end{document}